*Panda_NuPECC05, submitted to Nucl. Phys. News, 20 December 2005*# Exploring the Mysteries of Strong Interactions –

# The PANDA Experiment

KAI-THOMAS BRINKMANN[1], PAOLA GIANOTTI[2], AND INTI LEHMANN[3]
[1]*Institut für Kern- und Teilchenphysik, Technische Universität Dresden, D-01062 Dresden, Germany*
[2]*INFN-Laboratori Nazionali di Frascati, Via E. Fermi 40, I-00044 Frascati, Italy*
[3]*Department of Radiation Sciences, Uppsala University, Box 535, SE-75121 Uppsala, Sweden*## Introduction

The physics of strong interactions is undoubtedly one of the most challenging areas of modern science. Quantum Chromo Dynamics (QCD) is reproducing the physics phenomena only at distances much shorter than the size of the nucleon, where perturbation theory can be used yielding results of high precision and predictive power. As the coupling constant rises steeply at nuclear scales (see Figure 1) perturbative expansions diverge and a different theoretical approach is required. However, the strong interaction keeps providing new experimental observations, which were not predicted by "effective" theories. The latter retain the fundamental symmetries of QCD, but have problems in describing all the observed phenomena simultaneously.

The physics of strange and charmed quarks holds the potential to connect the two different energy domains interpolating between the limiting scales of QCD. In this regime only scarce experimental data is available, most of which has been obtained with electromagnetic probes.

One possible single issue that may greatly advance our understanding of hadronic structure is the predicted existence of states outside of the two- and three-quark classifications, which for example could arise from the excitation of gluonic degrees of freedom. Recent findings from running experiments at B-factories [2] show that, indeed, unexpected narrow states unaccounted for in the naïve quark models exist. Experiments focussed on the abundant production and systematic studies of these states are needed. Preferably, these should be performed using hadronic probes because the cross sections are expected to be very large in such systems. Results of high statistical precision are a decisive element to be able to identify and extract features of these exotic states. Hadronic probes are advantageous also for the production of hadrons with non-exotic quantum numbers, as these can be formed directly with high cross sections. Phase space cooling of the antiproton beam furthermore allows high precision determination of the mass and width of such states. Using heavier nuclei as targets enables us to investigate in-medium properties of hadrons and to produce hypernuclei, even those containing more than one strange quark, copiously.

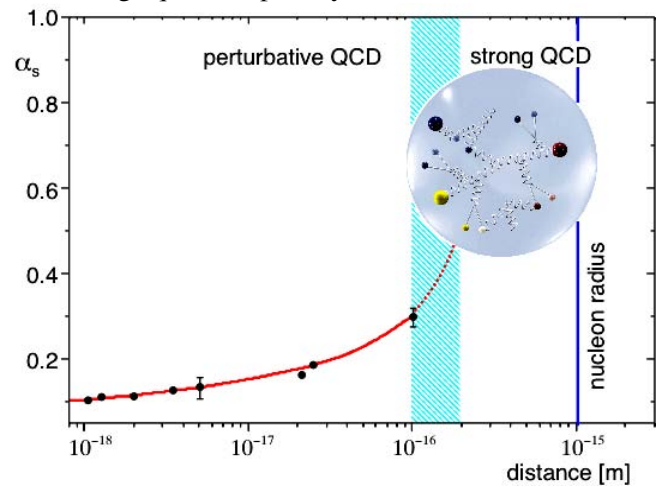

*Figure 1. Coupling constant of the strong interaction as a function of distance. The data points represent experimental values [1]. For distances between quarks comparable to the nucleon size the interaction becomes so strong that quarks cannot be further separated (confinement) and hadrons are formed. PANDA will investigate the properties of the strong interaction in this key region for the understanding of matter.*

The PANDA (antiProton ANnihilation at DArmstadt) experiment (see Ref. [3] for a more detailed description), which will be installed at the High Energy Storage Ring for antiprotons of the upcoming Facility for Antiproton and Ion Research (FAIR), features a scientific programme devoted to:
- charmonium spectroscopy,
- gluonic excitations (hybrids, glueballs),
- open and hidden charm in nuclei,
- $\gamma$-ray spectroscopy of hypernuclei,

and selected other topics, which will be studied with unprecedented accuracy.

## Burning Physics Questions

"Exotic objects" which, in their quantum numbers, cannot be described by configurations of two or three quarks, are an important consequence of

QCD. Such states would either include additional valence quarks or valence-glue content. To date only hints of such states were observed, but unambiguous evidence is missing. To a large extent, our understanding of hadron structure is based on the constituent quark model where mesons are quark-antiquark pairs, and baryons are made of three quarks. This may change if, indeed, other configurations were proven to exist.

The relation between the world of hadrons and the world of quarks (and gluons) is not fully understood. Only a few percent of the mass and part of the spin of a nucleon can be linked to valence quark contributions. The rest has to be attributed to a dynamical generation process which, to date, cannot be predicted concurrently by model calculations.

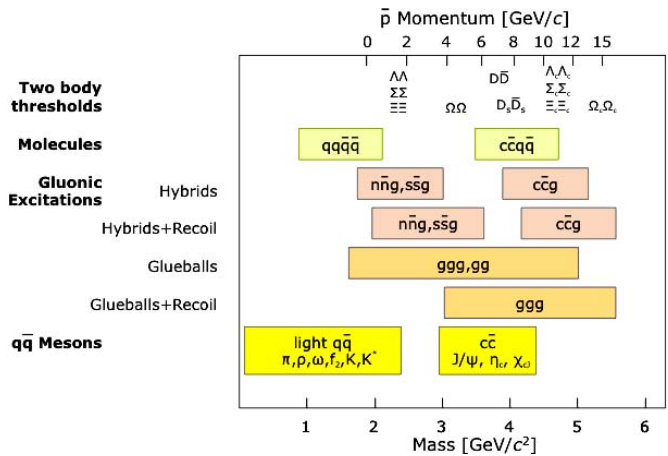

*Figure 2. Mass range of hadrons that will be accessible at PANDA. The upper scale indicates the corresponding antiproton momenta required in a fixed-target experiment. The HESR will provide 1.5 to 15 GeV/c antiprotons, which will allow charmonium spectroscopy, the search for charmed hybrids and glueballs, the production of D meson pairs and the production of Σ baryon pairs for hypernuclear studies.*

Chiral symmetry is one of the fundamental symmetries of QCD in the limit of massless quarks. This symmetry has to be broken to generate the hadrons, but it is predicted to be partially restored in nuclear matter [4]. To investigate the modification of hadron properties, when those are produced inside the nuclear medium, is of paramount importance to understand this mechanism. Some experimental evidence of mass shifts of light mesons has already been observed. It is, however, of utter importance to extend these studies to charmed mesons, as only then a complete picture could be drawn.

The strong interaction can be studied by replacing an up or down quark inside a nucleus by a strange quark. This introduces a new flavour, leading to the formation of a hypernucleus. Antiproton beams at FAIR will allow to efficiently produce hypernuclei even with more than one strange hadron, opening new perspectives for nuclear structure studies. The accessible set of hyperon states is not restricted in the population of nuclear states. This offers new and exciting perspectives for nuclear spectroscopy and the possibility to study the forces between hyperons and nucleons.

*Experimental Approach*

Conventional as well as exotic hadrons can be produced by a range of different experimental means. Among these, hadronic annihilation processes, and in particular antiproton-nucleon and antiproton-nucleus annihilations, have proven to possess all the necessary ingredients for fruitful harvests in the hadron field.

- Hadron annihilations produce a gluon-rich environment, a fundamental prerequisite to copiously produce gluonic excitations.
- The use of antiprotons permits to directly form all states with non-exotic quantum numbers (formation experiments). Ambiguities in the reconstruction are reduced and cross sections are considerably higher compared to producing additional particles in the final state (production experiments). The appearance of states in production but not in formation is a clear sign of exotic physics.
- Narrow resonances, such as charmonium states, can be scanned with high precision in formation experiments using the small energy spread available with antiproton beams (cooled to $\Delta p/p \approx 10^{-5}$).
- Since exotic systems will appear only in production experiments the physics analysis of Dalitz plots becomes important. This requires high-statistics data samples. Thus, high luminosity is a key requirement. This can be achieved using an internal target of high density, large numbers of projectiles and a high count-rate capability of the detector. The latter is mandatory since the overall cross sections of hadronic reactions are large while the cross sections of reaction channels of interest may be quite small.
- As reaction products are peaked around angles of 0° a fixed-target experiment with a magnetic spectrometer is the ideal tool. The physics topics as summarised in Figure 2 imply that the momentum range of the antiproton beam should extend up to 15 GeV/c with luminosities in the order of $10^{32}$ $cm^{-2}s^{-1}$.

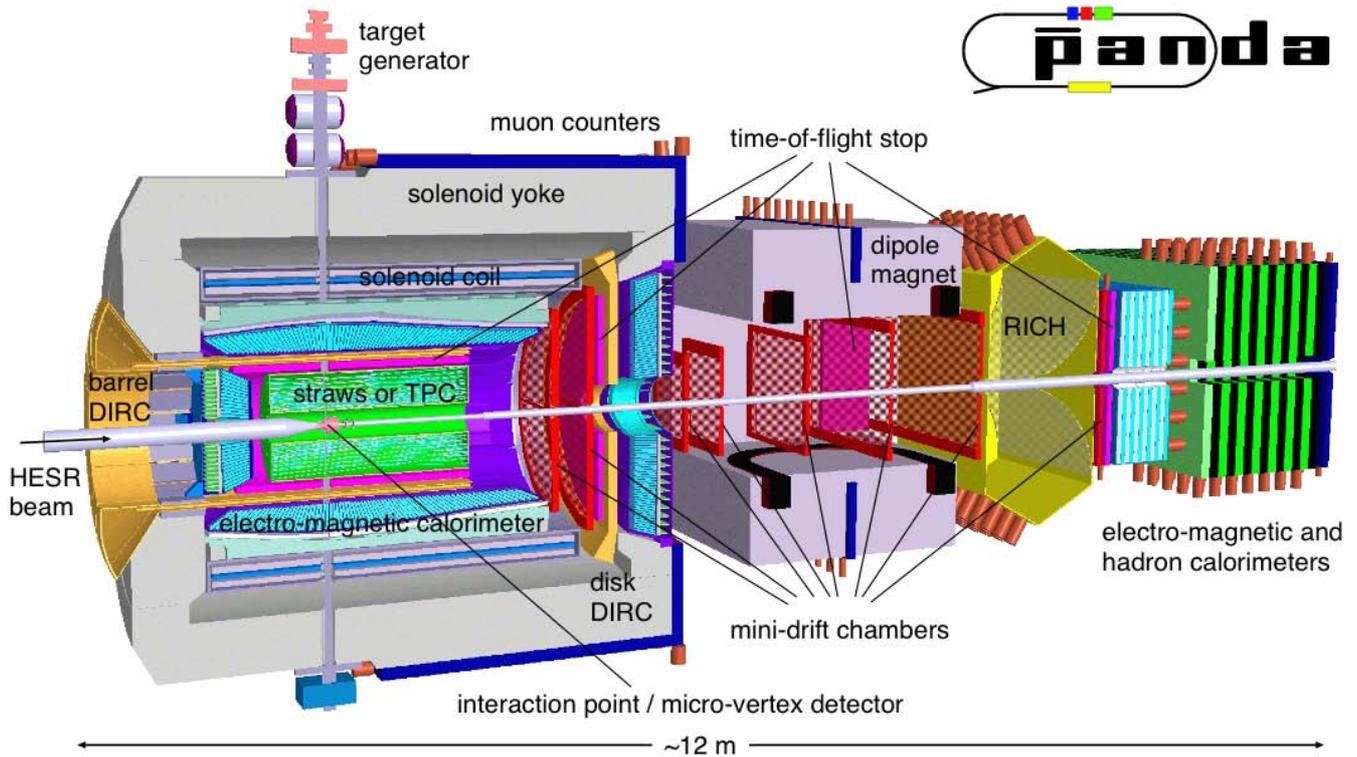

*Figure 3. Sketch of the PANDA detector. Antiprotons from the HESR enter from the left and meet clusters or pellets of hydrogen at the interaction point, which is surrounded by detectors contained in a solenoid. Angles up to 10° are covered by a magnetic forward spectrometer. For hypernuclear studies the beam up-stream endcap of the solenoid will be replaced with an array of Ge detectors and a dedicated target system. Details are described in the text and Ref [1].*

## The PANDA Apparatus

Following the above considerations, the PANDA detector has been designed as a $4\pi$ internal fixed-target detector at the High-Energy Storage Ring (HESR) for antiprotons, which is part of the FAIR facility.

29 GeV protons from the SIS 100 will be used to produce antiprotons, which are collected and cooled, and then injected into the HESR, where they are accelerated, cooled and stored at momenta between 1.5 and 15 GeV/c. In normal operation, the HESR will hold $10^{11}$ circulating antiprotons. At a length of about 500 m, the HESR will be equipped with an electron cooler as well as stochastic cooling to counterbalance the beam heating due to the internal target of PANDA. Stable operation will be accomplished with luminosities of $2\times10^{32}$ cm$^{-2}$s$^{-1}$ at relative momentum spreads of about $10^{-4}$. For experiments that need optimum momentum resolution, a second mode is foreseen where lower beam currents allow relative momentum spreads of about $10^{-5}$ at luminosities of $2\times10^{31}$ cm$^{-2}$s$^{-1}$. Details on the FAIR accelerator complex can be found in a separate contribution to this journal.

Figure 3 shows a sketch of the multi-purpose detector PANDA. Its magnet layout will combine a forward dipole spectrometer with a solenoid field region around the target. Despite the strong forward boost of a fixed-target experiment, coverage in both forward and backward direction is crucial for reactions where low-energy pions, kaons, and photons are emitted in $4\pi$.

Angles below 5° and 10°, in vertical and horizontal direction respectively, will be covered by a 2 Tm dipole spectrometer with a large gap, which will be equipped with drift chamber tracking, particle-identification layers as well as electromagnetic and hadronic calorimeters.

The central superconducting solenoid surrounding the target has a maximum field of 2 T. Photons and electrons will be measured in an electromagnetic calorimeter of high granularity. Due to the operation inside the coil and the instrumented flux return yoke, avalanche-photo diode readout is foreseen for the 19,000 crystals of the calorimeter. A Cherenkov detector of DIRC type [5] in the barrel structure, augmented by disk DIRC and a forward ring-imaging Cherenkov detector (RICH), will allow efficient particle identification down to momenta of about 750 MeV/c.

For the outer part of the tracking section a barrel of straw detectors is being designed. Alternatively, a time projection chamber (TPC) is considered as a challenging but advantageous option. The lightweight design of the tracking section and the self-supporting structure is chosen to minimize straggling and photon conversion. A micro-vertex detector will be operated inside the tracking detector. In order to get an accurate image of the particle tracks, its innermost active layer will be placed at a distance of 1 cm from the interaction point only.

Experiments on the proton require a target, which provides densities of hydrogen in the order of several times $10^{15}$ atoms/cm$^2$ at the interaction point without disturbing the ultra-high vacuum of the storage ring. This is especially challenging, as detectors are to cover a large solid angle very close to the interaction point. Thus, feed tubes and vacuum pipes may occupy only very limited space. Both cluster-jet and pellet targets are under development and seem capable to fulfil those conditions with advantages concerning homogeneity and vertex definition, respectively. For experiments on heavy ions, heavy gases may be used in these systems or fibre targets may be employed. An additional set-up (not shown in Figure 1) will allow high-resolution spectroscopy of excited hypernuclei at PANDA. It includes primary and secondary targets, an array of high-resolution germanium detectors, and a dedicated tracking system, which will replace the upstream endcap of the calorimeter and the micro-vertex detector.

## *Summary*

Our knowledge of the behaviour of QCD at large distances is still rather scarce. Spectroscopy experiments within hadron physics are the tool to investigate both the dynamics governing the interaction of fundamental particles and the existence of new forms of matter. The latter could consist of states with gluonic degrees of freedom, such as glueballs and hybrids, previously undiscovered charmonium states, the extension of the nuclear chart into the strangeness dimension, or particles produced inside nuclear matter. Only if we are able to predict, confirm, and explain the physical states of the theory, we may claim to understand the strong interaction. The deep understanding might have further far-reaching implications in nuclear and particle physics.

The PANDA collaboration is prepared to address these topics with a general-purpose internal-target experiment utilising the antiprotons provided at the upcoming FAIR facility. Within the growing PANDA collaboration of currently 350 physicists from 15 countries, an extensive R&D programme is under way, which comprises already a detailed design of the detector. PANDA gratefully acknowledges the support of the respective national research agencies and the European Union funds (see Ref. [3]).

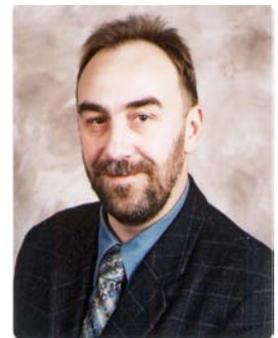

KAI-THOMAS BRINKMANN
*TU Dresden*

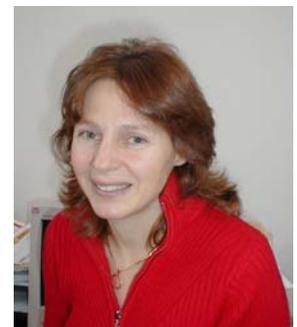

PAOLA GIANOTTI
*INFN Frascati*

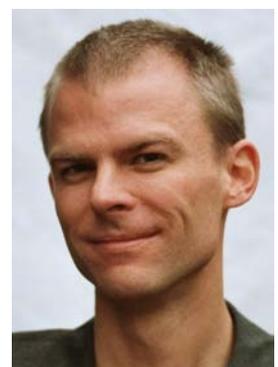

INTI LEHMANN
*Uppsala University*